\documentclass[11pt,twoside]{article}
\usepackage{cozumel2005}
\usepackage{epsf}
\usepackage{psfig}
\usepackage{lscape}
\usepackage{graphicx}
\begin{document}
\title{Red clump distances to the inner Galactic structures}    
\author{C. Babusiaux$^{2,1}$ and G. Gilmore$^1$}   
\affil{$^1$ Institute of Astronomy, University of Cambridge, Cambridge, CB30HA, UK\\
$^2$ Institut d'Astronomie et d'Astrophysique, Universit\'e Libre de Bruxelles, 
B-1050 Bruxelles}    

\begin{abstract} 
The least well known structure of the Galaxy is its central region, 
because of the high extinction, the crowding and the confusion between 
disk and bulge sources along the line of sight.
We show how the structure of the inner Galaxy and the dust 
distribution can be strongly constrained by using red clump stars as 
distance indicator.
The results of this method applied on the 
deep near-infrared survey of the inner Galactic bulge 
made with the Cambridge Infrared Survey Instrument (CIRSI) are presented. 
\end{abstract}


\section{Introduction}                     

The study of the Milky Way Galaxy offers a unique opportunity for understanding the complexity of galaxy formation and evolution. However our point of view from within the Galactic disk has its drawbacks when trying to obtain large scale properties of the Galaxy, and in particular of its central region. The high interstellar extinction, the crowding and the confusion between foreground disk stars and bulge sources complicates the studies of the inner Galactic regions. 
We would like to know how the bulge formed and evolved, what is it link with the disk and the halo. But we still doesn't know the shape of the bulge. We still can't say what the Milky Way looks like from the outside. 

The central region of the Galaxy has been subject to many studies in the past few years, thanks to technology developments allowing wider and deeper multi-wavelength surveys. There is now substantial evidence for the presence of a triaxial structure in the inner Galaxy. Those evidences mainly come from gas kinematics, infrared luminosity distribution, star counts, and since recently, red clump stars. 
There is general agreement that the near end of this triaxility is in the first quadrant, however the values given for its orientation with respect to the Sun-Galactic center line range from 10$\deg$ to 45$\deg$ and can be interpreted either as a triaxial bulge or a bulge plus a bar. 

To disentangle the different structures that may be present in the inner Galaxy, distance estimators are needed. Red clump stars are well known distance indicators thanks to their bright and narrow luminosity distribution. They were first used to constrain the Galactic bar structure by the OGLE team (\citealt{STA94}, \citealt{STA97}), followed very recently by \cite{BAB05} and \cite{NIS05}. We present here this method with its application to the CIRSI Galactic bulge survey of \cite{BAB05}.

\section{The CIRSI Galactic bulge survey}

The CIRSI Galactic bulge survey is a deep near-infrared survey of the bulge/bar in the Galactic plane. Fields have been selected at Galactic latitudes $-0.3\deg<b<0.3\deg$ and Galactic longitudes $\ell=\pm5.7\deg$ and $\ell=\pm9.7\deg$. A field at $\ell=0\deg$, $b=1\deg$ has also been included as a calibration field of the expected asymmetry caused by the Galactic bar between positive and negative longitudes. 
The observations in the J, H and K$_s$ were obtained with the Cambridge Infrared Survey Instrument (\citealt{BEC97}, \citealt{MAC00}) on the du Pont 2.5m telescope at Las Campanas Observatory.   CIRSI is a mosaic imager with a pixel scale of 0.2 arcsec pixel$^{-1}$, leading to a field of view of about 13 x 13 arcmin$^2$. The magnitude limits are around J=20, H=18.5 and K$_s$=18 mag. 
The details of the observations and data reductions are presented in \cite{BAB05}.

\section{Deriving the distances from red clump stars}

\subsection{The method}


To counteract the effect of extinction, reddening-independent magnitudes can be derived using any two colors for a given extinction law, for example: 

\begin{equation}
{K_s}_{J-K_s} = K_s - {A_{K_s} \over A_J-A_{K_s}}(J-K_s)
\label{eqme}
\end{equation}

In the near-infrared, the extinction curve is well defined by a single power-law function \citep{CAR89}. Unlike in optical studies where extinction law variations remain a source of possible systematic uncertainty (\citealt{UDA03}, \citealt{SUM04}), near-infrared  reddening-independent magnitudes should not be significantly affected by variations of the interstellar medium properties along the different lines of sight. In the following we adopt $A_{K_s}/E_{J-K_s}$ = 0.64 from \cite{HE95}.

\begin{figure}[t]
\centering
\includegraphics[angle=-90,width=0.7\textwidth]{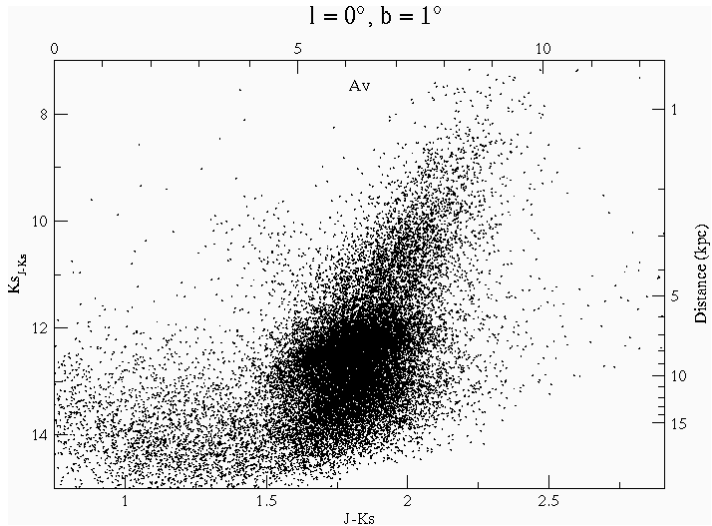}
\caption{Color magnitude diagram of the bulge calibration field 
using reddening independent magnitudes. 
In this diagram, increasing extinction moves a star
horizontally to the right, while increasing distance moves it
vertically downwards. Locations of red clump stars on this diagram as
a function of absorption and distance are given by the top and right
hand axes.}
\label{fig:CMDC32}
\end{figure}

In a color magnitude diagram (CMD) made with a reddening-independent
magnitude, the effect of distance and extinction are disentangled: a
star of a given spectral type moves vertically with distance,
horizontally with extinction. This is illustrated in figure \ref{fig:CMDC32}, which is the color - reddening-independent magnitude diagram of the low-extinction minor axis bulge field at ($\ell=0\deg$, $b=1\deg$).
On this diagram the bulge red giant branch from a well-defined, almost single distance feature with a distinct red clump. The location of the red clump
stars in this ($J-K_s$, ${K_s}_{J-K_s}$) diagram is translated into an ($A_V$, distance) estimate on the opposite axes of figure \ref{fig:CMDC32}. 

To quantify the distance estimate given by the red clump stars from CMDs like figure \ref{fig:CMDC32}, a non-linear least-squares fit of function \ref{eqfit} to the histogram of reddening-independent magnitudes of red giant stars is applied, following \cite{STA98} and \cite{SAL02}:

\begin{equation}
N(m) = a + b m + c m^2 + {N_{RC} \over \sigma_{RC} \sqrt{2\pi}}
\exp[-{(m_{RC}-m)^2 \over 2 \sigma_{RC}^2}]
\label{eqfit}  
\end{equation}

\begin{figure}[t]
\centering
\includegraphics[width=0.7\textwidth]{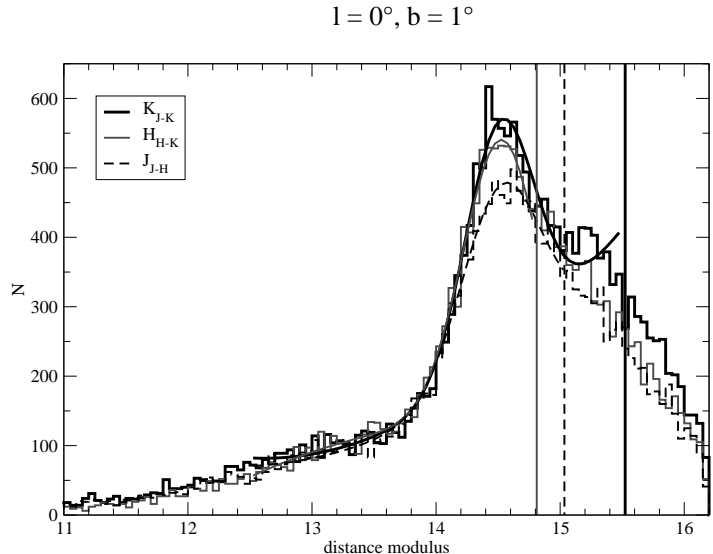}
\caption{The distance modulus distribution computed for red clump stars using reddening-independent magnitudes. This is for the bulge calibration field, using alpha-enhanced colors. The vertical lines show estimates of the photometric completeness limits. Fits to equation \ref{eqfit} are overlaid.}
\label{fig:histoC32}
\end{figure}

The Gaussian term represents a fit to the bulge/bar red clump.  The
first three terms describe a fit to the background distribution of
non-bulge-clump red giant stars, which in our case contains not only
the other bulge giants, but also the disk red clump stars that fall
into the bulge red clump. This fit is illustrated in figure \ref{fig:histoC32}.

\subsection{Calibration of the distances}

To translate reddening-independent magnitudes into distance modulus estimates, the red clump absolute magnitude and colors need to be known. 
We used the red clump stars colors predicted by the Padova isochrones provided in the 2MASS system \cite{BON04} for a 10 Gyr population of solar metallicity: ($J-K_s$)$_0$=0.68 and ($H-K_s$)$_0$=0.07.
We calibrated the absolute magnitude of the red clump ${M_0}_{K_s}$ using our bulge calibration field at ($\ell=0\deg$, $b=1\deg$).

If we assume the absolute red clump distance of $M_K=-1.61 \pm 0.03$ mag, 
derived by \cite{ALV00} for the Hipparcos red clump, 
we obtain a distance for the Galactic center of $D_{GC}=7.6 \pm 0.15$ kpc.
However according to \cite{SAL02}, a small population correction 
should be applied to the \cite{ALV00} calibration for bulge stars. 
Using their population correction for Baade's window with solar metallicity
we derive $M_K=-1.68$ mag and $D_{GC}=7.8$ kpc, while
with enhanced $\alpha$-elements it leads to $M_K=-1.72$ mag 
and $D_{GC}=8.0$ kpc. However if we assume $\alpha$-enhancement, 
we also have to change the assumed ($J-K_s$)$_0$, which leads us back to $D_{GC}=7.7 \pm 0.15$ kpc. All those values are consistent with the latest Galactic center distance estimates (e.g. \citealt{REI93}, \citealt{MCN00}, \citealt{EIS03}) which give $D_{GC} = 8 \pm 0.5$ kpc. 
Considering the different uncertainties in the absolute magnitude of the 
red clump, we decided to calibrate this latter to $M_{K_s}^{RC}=-1.72$ mag, 
assuming a distance for the Galactic center of 8 kpc. 

As we calibrated the absolute magnitude of the red clump on the bulge field at $\ell=0\deg$, our distance estimates should not depend on the population corrections suggested by \cite{SAL02}. However this applies only if we assume that we are probing the same population in the different fields of the survey. A hint that this may not be the case is given by the fact that the fits of the reddening-independent magnitudes derived with our different colors do not perfectly match for the calibration field at $\ell=0\deg$ if we assume colors derived for solar abundances, while the fits converge if we assume colors derived for $\alpha$-enhanced abundances (figure \ref{fig:histoC32}). On the opposite the fits match better for solar abundance colors for the other fields of the survey (figure 5 of \citealt{BAB05}).

\begin{figure}[t]
\centering
\includegraphics[width=0.7\textwidth]{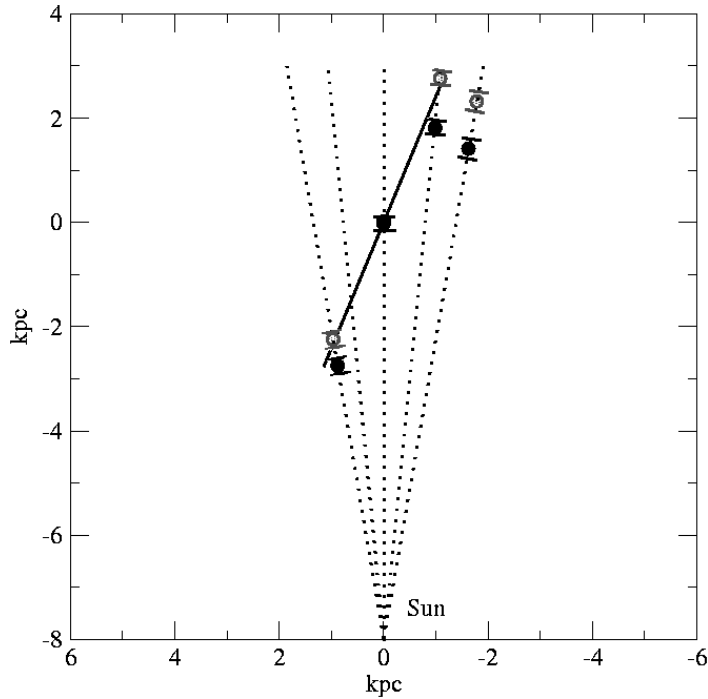}
\caption{Red clump distances projected on the Galactic plane seen from the North Galactic pole. The dotted lines indicate the line of sights of this study:
$\ell=\pm9.7\deg$, $\ell=\pm5.7\deg$ and $\ell=0\deg$. The solid circles are the distances derived using solar abundances for all directions. The empty circles are derived assuming for the bulge field at $\ell=0\deg$ a 10 Gyr old population with enhanced $\alpha$-element abundances, while 
assuming for all the other fields a 6 Gyr old population of solar abundances.
The black line illustrates a Galactic bar of 3~kpc radius inclined
at 22.5\deg\ to the Sun-Galactic Center line.}
\label{fig:distshifts}
\end{figure}

To illustrate the effect of systematics in the assumed stellar populations, figure \ref{fig:distshifts} presents the distances estimated from the red clump stars with two different hypothesis. First we assume the same calibrations for all fields. Then we assume for the calibration bulge at $\ell=0\deg$ a 10 Gyr old population with enhanced $\alpha$-element abundances (\citealt{MCW94}, \citealt{MAT99}) and for the others, associated with a younger bar population, an age of 6 Gyr \citep{COL02} and solar abundances. The deduced distances of the second hypothesis show a better fit to a simple linear regression model for the Galactic bar structure. If confirmed by spectral observations, this implies a bar population which is formed from the inner disk and not from the old bulge.

\section{Constraining the structure of the inner Galaxy}

\subsection{The asymmetric structure}

\begin{figure}[t]
\centering
\includegraphics[angle=-90,width=0.45\textwidth]{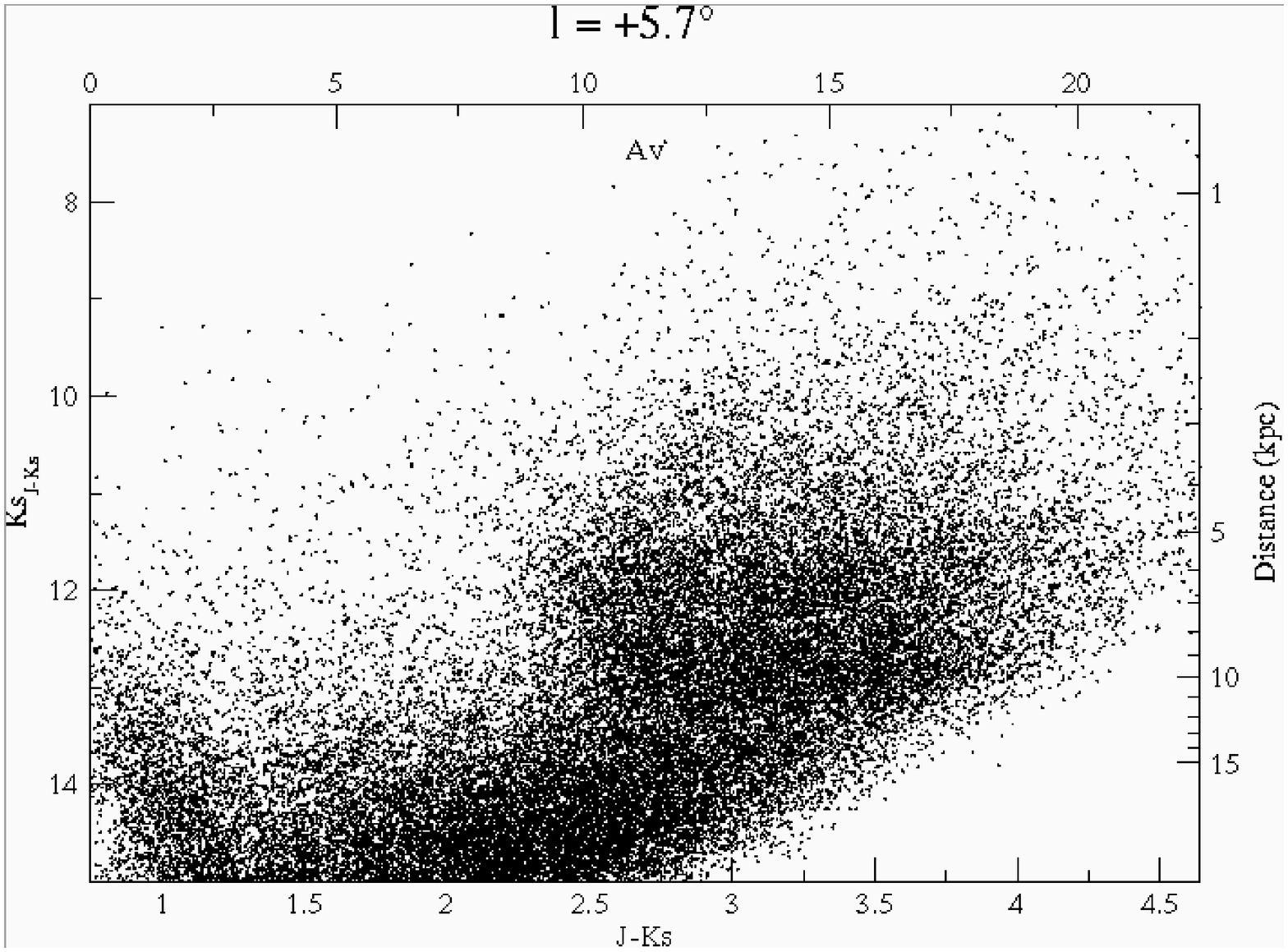}
\hspace{0.05\textwidth}
\includegraphics[angle=-90,width=0.45\textwidth]{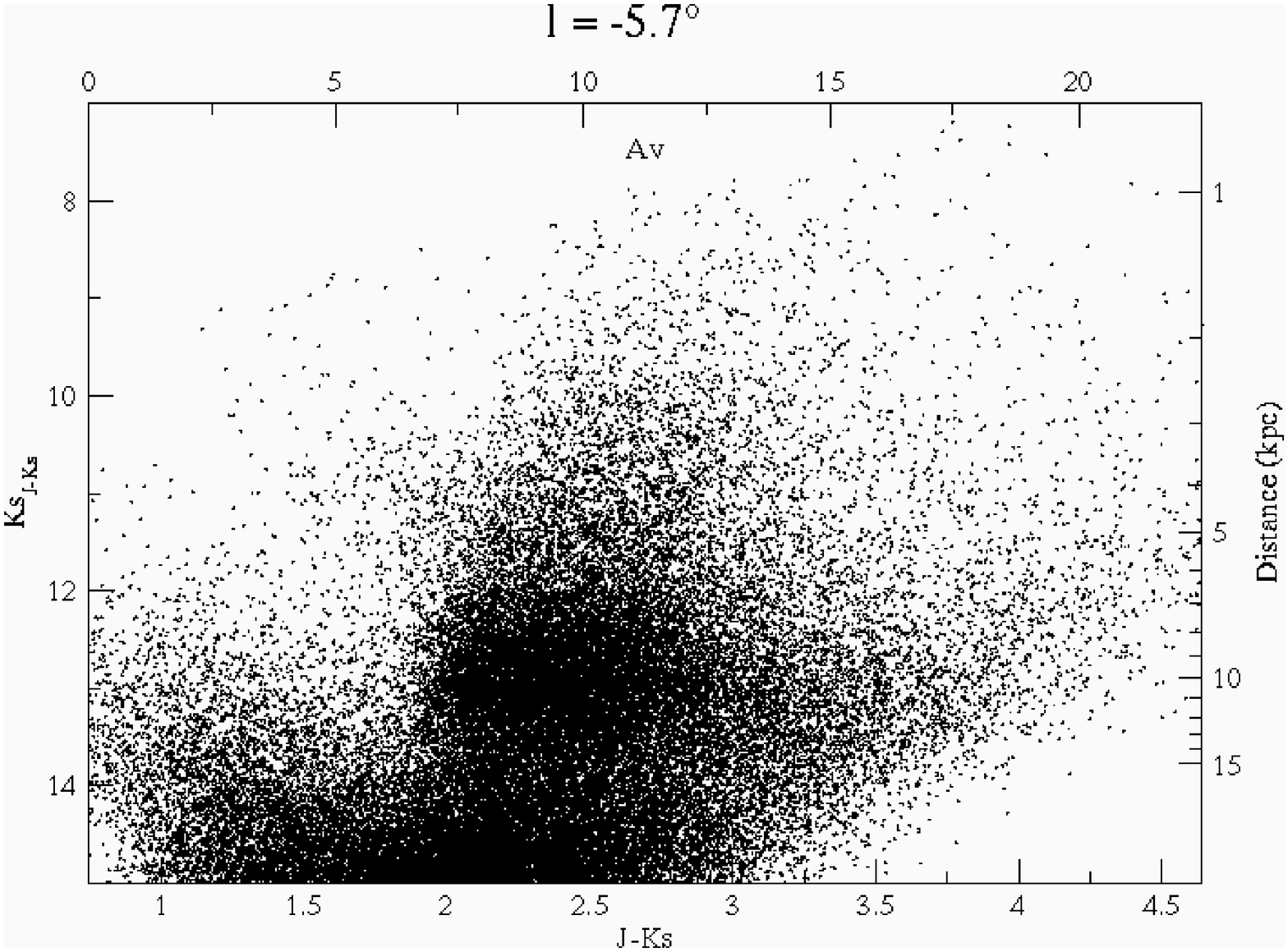}
\par \vspace{3mm}
\includegraphics[angle=-90,width=0.45\textwidth]{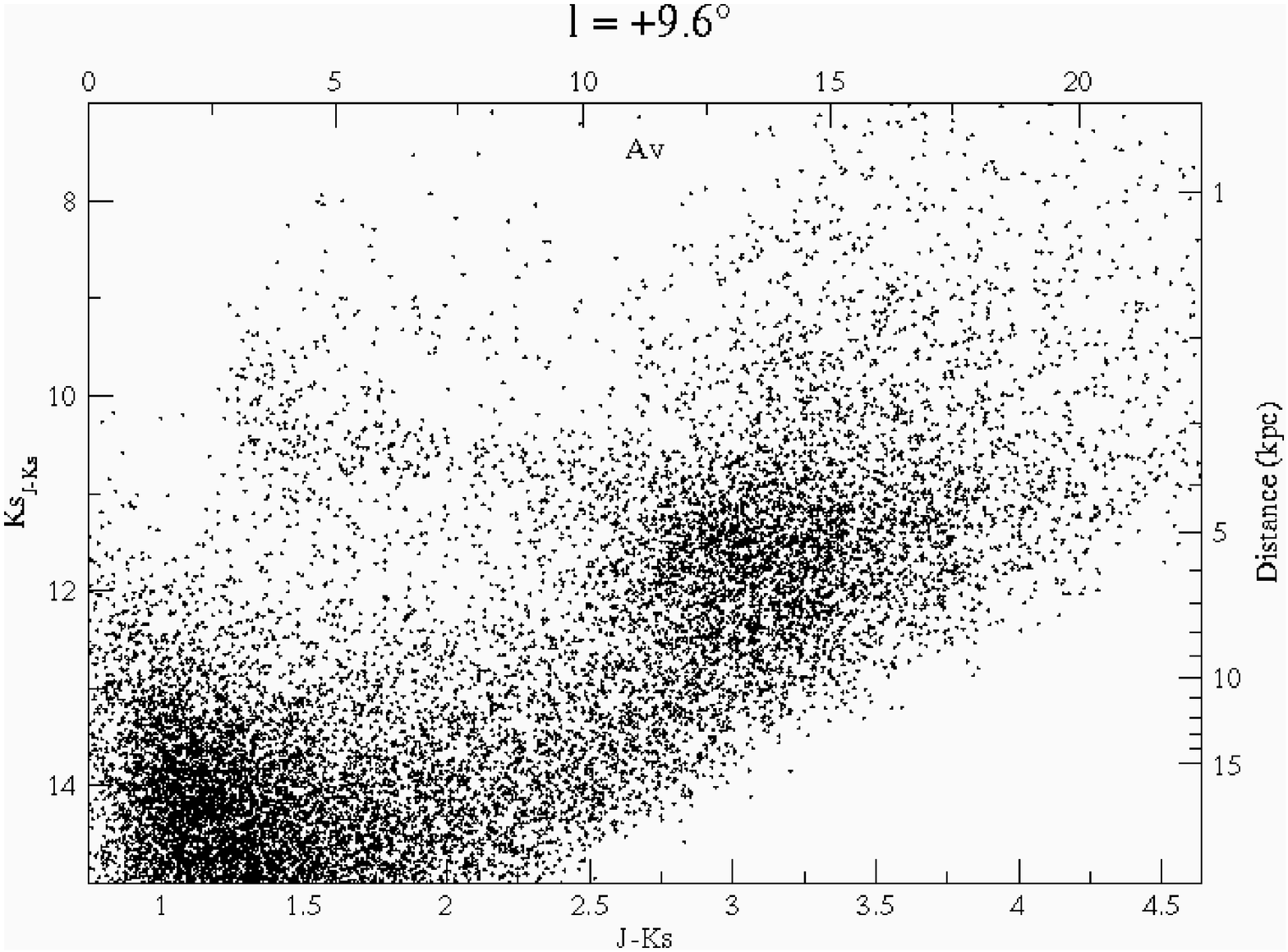}
\hspace{0.05\textwidth}
\includegraphics[angle=-90,width=0.45\textwidth]{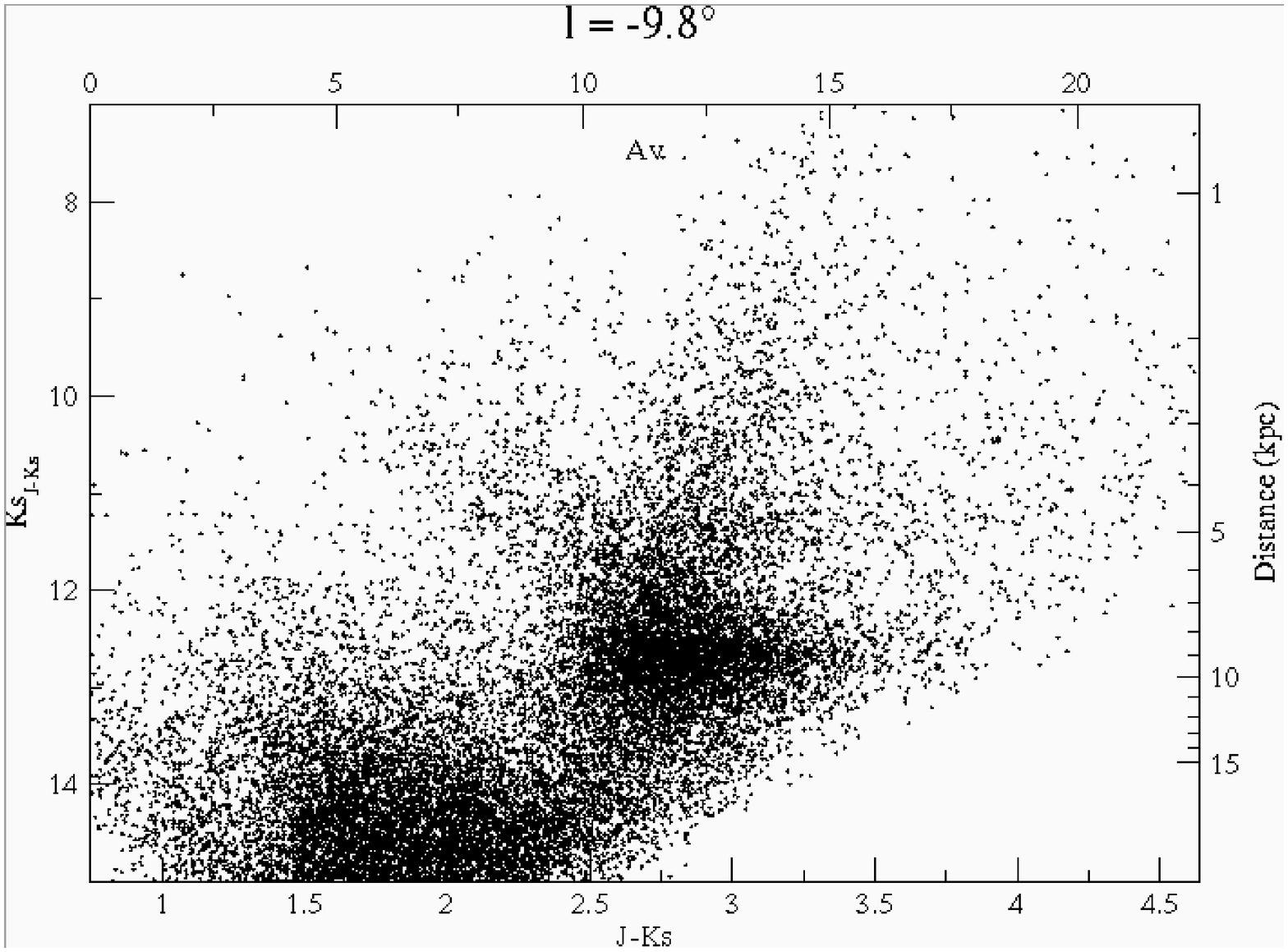}
\caption{Same as figure \ref{fig:CMDC32} for the four pointing directions at b=0$\deg$, l=$\pm5.7\deg$ and l=$\pm9.7\deg$.}
\label{fig:CMDs}
\end{figure}

\begin{figure}[t]
\centering
\includegraphics[width=0.7\textwidth]{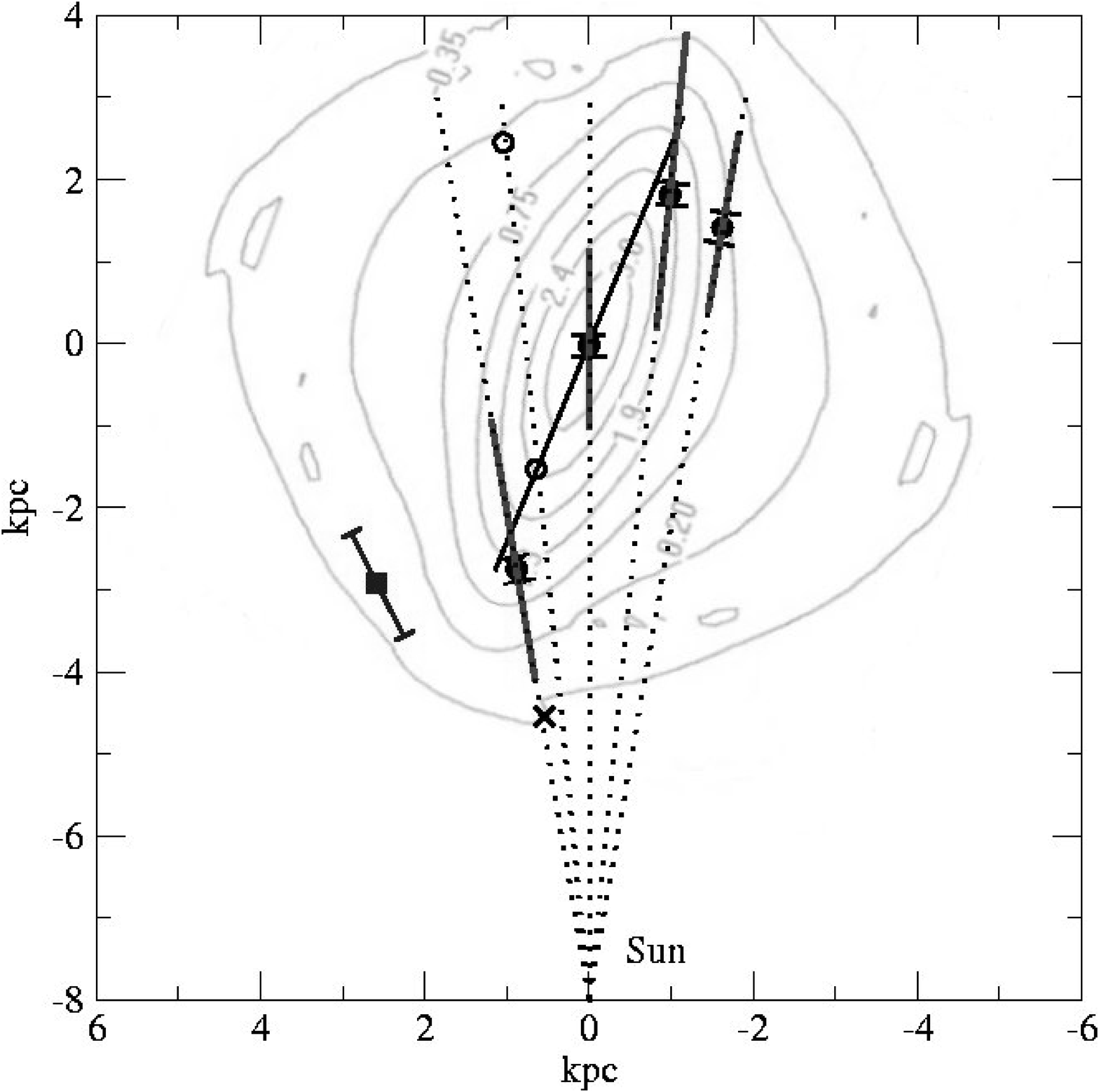}
\caption{Same as figure \ref{fig:distshifts}. 
Thick grey lines along the lines of
sight through each mean distance represent the one$-\sigma$ range in
distances deduced from a Gaussian fit to the red clump apparent
distance modulus distribution. The background contour
map is a plane projection of the Bissantz \& Gerhard (2002) Galactic
bulge bar model, from their figure 11.  The empty circles at
$\ell=+5.7\deg$ are only indicative, as the more distant point may be
confused by background disk stars, as discussed in the text.  The small
cross at 3.5kpc from the Sun along the direction $\ell=+9.6\deg$
indicates the position of a local high-extinction region detected in
our photometry. The square to the left of the dotted lines through our
data indicates the mean position of the red clump detected by
Hammersley et al. (2000) at $\ell$=+27\deg.}
\label{fig:barmod}
\end{figure}

The color reddening-independent magnitude diagrams of our different survey fields show clear differences in the distribution of their stellar densities (figure \ref{fig:CMDs}). The derived spatial map is summarized in figure \ref{fig:barmod}. The mean distances down each line of sight are represented as well as an indication of the distance range. This distance range corresponds to the measured 1$\sigma$ dispersion in distance modulus deconvolved from the intrinsic dispersion of the red clump luminosity and the photometric errors. This is only a qualitative estimation as the distribution in distance has no reason to be Gaussian. Furthermore, due to our observation angle our distance estimates are slightly biased towards larger distances. On the other hand those distance estimates are independent of any Galactic model. 

The mean red clump distances measured at $\ell=+9\deg$ and
$\ell=-5\deg$ are consistent with the more recent bar models,
particularly those derived considering both gas dynamics and
available surface brightness luminosity distributions, and the model derived 
from the OGLE data by \cite{STA97}.
Those models deduce a bar orientation with respect to the Sun-Galactic center
direction of $15\deg \la \phi_{bar} \la 35\deg$
(e.g. \citealt{GER01}, \citealt{MER04}). The model of \cite{BIS02}
(their figure 11) is overlaid on figure \ref{fig:barmod}. This model
assumes a bar with an angle $\phi_{bar}=20\deg$, a length of 3.5~kpc
and an axis ratio 10:3-4. Their is a good overall agreement with our direct
distance determinations.

Our red clump distance derived at $\ell=-9\deg$ is however not
simply consistent with the bar model described above. A single linear
fit to the four mean distances measured at $\ell=+9,0,-5,-9\deg$
has all four points more than 3$\sigma$ from the `best fit'.  
\cite{NIS05} also find that a simple linear fit does not reproduce their data 
but their distance estimates are not coherent with ours and they still find a highly significant distance difference between $\ell=-5\deg$ and $\ell=-9\deg$. 
To understand the 
discrepancy with \cite{NIS05} examinations of both CMDs are needed. 
The OGLE data also seem to infer that a flattening of the distances occurs 
around $\ell=-5\deg$ (figure 6 of \citealt{SUM04}).
Our feature observed at $\ell=-9\deg$ does seem real: inspection of
figure \ref{fig:CMDs} shows a clear overdensity of the whole red giant
branch at the same distance as the mean clump distance, determined to
be $\sim$ 9~kpc, in this field. Moreover, in this line of sight, the
most likely source of distance bias, which is the presence of disk red clump
stars, would bias the distance estimate towards larger distances, not
shorter. On the opposite, this could have biased our distance estimate at $\ell=-5\deg$. 
A possible physical interpretation to the fact that our red clump distances 
measured at $\ell=-9\deg$ and $\ell=-5\deg$ are equivalent is
that we have detected the signature of the end of
the bar. Star count peaks have also been detected in this region in
studies of old OH/IR stars by \cite{SEV99} and in the DENIS star
counts by \cite{LOP01}. 

\cite{LOP01} suggest that their local density
maximum at $\ell=-9\deg$ may be related to the red clump overdensity detected by \cite{HAM00} at $\ell=27\deg$. The \cite{HAM00} density maximum is
indicated in figure \ref{fig:barmod} by a square.
 The presence of these two local features at $\ell=27\deg$ and $\ell=-9\deg$ 
and their mean red clump distance estimates
could imply the presence of a double triaxiality in the inner Galaxy, 
a double bar, or a triaxial bulge oriented at $\sim22\deg$ with a longer, 
thiner bar oriented at $\sim44\deg$. 
However, if such a second structure did exist, we should have detected 
its signature in our other survey fields. No such complex signature is evident
(cf. figure \ref{fig:CMDs}). 
If two complex spatial distributions were projected 
down our lines of sight without being resolved, this would imply that our 
distance determinations are biased so that 
the angle of the first structure is in fact smaller than the 22\deg\ we
measured, and that the spread of the red clump distances observed down
each line of sight would encompass both structures. The latter does
not seem to correspond to the rather small distance dispersions we
derived. Furthermore,
\cite{PIC03} do not detect a density excess at $\ell=+21\deg$, confirming
that the structure seen at $\ell=+27\deg$ could be local and unrelated to the 
larger bulge or bar. 

An alternative, and perhaps more consistent, interpretation of the
$\ell=-9\deg$ structure we observe is the
presence of a stellar ring or pseudo-ring at the end of the Galactic
bar. If so, the bar and the stellar ring would have a radius of
2.3$\pm$0.25~kpc. Our observations  at $\ell=+9\deg$ indicate that 
the radius of
the bar is at least 2.7$\pm$0.2~kpc long. Those two determinations are
consistent within 1$\sigma$. We note that assuming the bulge/bar stellar 
population differences used for figure \ref{fig:distshifts} 
leads to a larger bar radius of about 3 kpc.
A bar radius of about 2.5~kpc would agree
with the model of \cite{LEP00} and \cite{SEV99}, but is smaller than
the value indicated by \cite{GER01} in his review.  The presence of
a ring has been suspected by several authors.  A molecular ring at
about 4-5~kpc is a well known feature deduced from CO maps. 
\cite{COM96} link their derived distribution of
ultracompact HII regions to the molecular ring, but also detect a star
forming ring at about 2~kpc from the Galactic centre.  A stellar ring
was suspected to lie at 3.5~kpc from the Galactic centre by
\cite{BER95}.  From an OH/IR star study \cite{SEV99} suggests that an
inner ring lies between between 2.2 and 3.5~kpc. From star counts in
the DENIS survey \cite{LOP01} also argue for the presence of a stellar
ring, mainly from the detection of a density peak at $\ell=-22\deg$ that,
following \cite{SEV99}, they associate with the tangential point to
the 3-kpc arm, which is likely to be a (pseudo-)ring. The presence of
a Galactic ring would also help to reproduce the observed microlensing
optical depth (\citealt{SEV01}). 

\subsection{The line of sight towards $\ell=+5\deg$}

The reddening-independent colour-magnitude diagram for the line of
sight towards $\ell=+5\deg$ (figure \ref{fig:CMDs}) differs significantly
from those in other directions, in that the red clump stars indicate a
very considerable range of distances, with no clear local maximum
density. The broad apparent distance distribution observed for the
red clump stars at $\ell=+5\deg$ could be explained by the presence
of two features, one at $\sim$ 6~kpc and a second at $\sim$
11~kpc. The first distance range is consistent with the bar structure
confirmed above. It is also consistent with the fact that the OGLE
data of \cite{STA94}, observed at a lower latitude of $b=-3.5\deg$, 
show similar
photometric behaviour at $\ell=+5\deg$ compared to $\ell=-5\deg$,
except for the expected shift in distance modulus. 

The fainter feature in our low-latitude data could then be due to
another structure, further away. This feature could be due to disk red
clump stars. Indeed with increasing distance both the volume observed
and the extinction increase, and considering the logarithmic relation
between distance and magnitude, any distant red clump stars visible
would concentrate in the high-extinction high distance-modulus
bottom-right part of the CMD. Deeper data than we have available are
then needed to determine if we are indeed seeing the distant disk
beyond the bulge or a local structure which could be associated with the ring. 
This region is also interesting 
for further study as a broad velocity-width molecular clump has been 
detected at
$5.2\deg<\ell<6.0\deg$ (e.g. \citealt{BOY94}, \citealt{BIT97},
\citealt{DAM01}), which may be a manifestation of gas shocks
(\citealt{KUM97}, \citealt{FUX99}). An estimate of the distance of
those two structures, if the fainter one is real and indeed a local
maximum, is presented in figure \ref{fig:barmod} as the open circles.

\subsection{The dust distribution}

The red clump star locus on the color reddening independent magnitude diagrams allows to quantify the distribution of the extinction along the line of sight. 
For example, at $\ell=+9\deg$ a clear flattening of the disk red clump 
distribution at ${K_s}_{J-K_s}
\approx 10.5$ indicates the presence of a specific source of
extinction located at about 3.5~kpc from the Sun, which increases
$A_V$ by about 7.5~mag in only 1.5~kpc. This feature is consistent
with the steep increase in absorption at 3.4-4~kpc seen at
$\ell$=12.9\deg\ by \cite{BER95}, that they associated with the
molecular ring. It is indicated by a cross in figure \ref{fig:barmod}. 

In all our Galactic plane fields, the bulk of bulge red 
clump stars becomes visible at $A_V \sim 10$ mag, and apparently
suffer a range of internal extinction of more than $\Delta
A_V\sim5$~mag.
\cite{LOP01} detect an asymmetry in the Galactic plane extinction distribution,
 with more extinction at negative than at positive longitudes. 
We do not confirm this result. The fact that we do not find any
asymmetry in the extinction of red clump stars between positive and negative
longitudes (figure \ref{fig:CMDs}), while there is an asymmetry in
distance, is certainly consistent with a minimum in the dust
distribution in this region. 
Interior of the bar outer radius a lower global 
density of stars and gas is often seen in observations of other galaxies.  
Several authors have suggested that such a stellar density decrease may 
be present in the Galactic disk (\citealt{BIN91}, \citealt{BER95}, 
\citealt{FRE98}, \citealt{LEP00}, \citealt{LOP04}). Our observations are 
consistent with such a decrease in the dust distribution of the inner disk. 

 Another interesting feature often observed in barred galaxies is the 
presence of dust lanes associated with the bar. From DIRBE surface 
brightness maps, \cite{CAL96} suggested the presence of such a thick dust lane, preceding the Galactic bar in its rotation. 
They estimate the extra absorption present at 
negative longitudes to be between 1 and 2.6 mag in K. The presence of such 
a feature is clearly ruled out by our observations.

\section{Conclusion}

Red clump stars are very powerfull distance indicators to the inner Galactic structures. Colour - reddening-independent magnitude diagrams have been shown to
disentangle the effects of distance and extinction, allowing a direct
conversion of red clump star photometry ($J-K$,$K_{J-K}$) into an
($A_V$,distance) estimate. The use of the near-infrared allows to observe red clump stars within the Galactic plane, where the extinction is too high for optical studies, and provides reddening-independent magnitudes that are not influenced by variations of the ISM properties between the different lines of sights. 
Warnings should however be raised for contamination by disk red clump stars and differencial effects of population corrections. 

This method have been applied to the CIRSI Galactic bulge survey and produced the following main results:

\begin{itemize}
\item[-] The presence of a triaxial structure at the centre of our Galaxy
is confirmed.  Its angle relative to the Sun-Galactic centre line is
$\phi_{bar}=22 \pm 5.5 $\deg. It extends to at least 2.5~kpc from
the Galactic Centre. A large axis ratio is excluded, but our data are consistent
with a 10:3-4 ratio. In particular the distance dispersion of the bulge along
the ($\ell=0\deg$, $b=1\deg$) line of sight is less than 1~kpc.
\item[-] A structure present at $\ell=-9.8\deg$ is not aligned with this
triaxiality. We suggest that the structure present at $\ell=-9.8\deg$ is
likely to be the signature of the end of the Galactic bar, which is
therefore circumscribed by an inner pseudo-ring.
\item[-] A decrease in the dust distribution inside the bar radius is
inferred from the extinction distribution in our fields.
\item[-] Our observations are not consistent with the existence of the
dust lane preceding the Galactic bar at negative longitudes suggested
by \cite{CAL96}.
\end{itemize}




\end{document}